\documentclass[conference]{IEEEtran}
\IEEEoverridecommandlockouts

\usepackage{enumitem}

\usepackage{indentfirst}
\usepackage{amsmath}
\usepackage{graphicx}
\usepackage{booktabs}
\usepackage{multirow}
\usepackage{booktabs, makecell, siunitx, tabularx}
\usepackage{cite}
\usepackage{amsmath,amssymb,amsfonts}
\usepackage{algorithmic}
\usepackage{graphicx}
\usepackage{textcomp}
\usepackage{xcolor}
\usepackage{hyphenat}
\def\BibTeX{{\rm B\kern-.05em{\sc i\kern-.025em b}\kern-.08em
    T\kern-.1667em\lower.7ex\hbox{E}\kern-.125emX}}
\begin{document}

\title{Neuro-MSBG: An End-to-End Neural Model for Hearing Loss Simulation\\
 }

\author{%
  \IEEEauthorblockN{%
    Hui\hyp Guan Yuan\IEEEauthorrefmark{1}\IEEEauthorrefmark{2},\ 
    Ryandhimas E.~Zezario\IEEEauthorrefmark{1},\ 
    Shafique Ahmed\IEEEauthorrefmark{1},\ 
    Hsin\hyp Min Wang\IEEEauthorrefmark{1},\ 
    Kai\hyp Lung Hua\IEEEauthorrefmark{2}\IEEEauthorrefmark{3},\ 
    Yu Tsao\IEEEauthorrefmark{1}%
  }
  \vspace{1ex}
  \IEEEauthorblockA{%
    \IEEEauthorrefmark{1}Academia Sinica, Taipei, Taiwan\\
    \IEEEauthorrefmark{2}National Taiwan University of Science and Technology, Taipei, Taiwan\\
    \IEEEauthorrefmark{3}Microsoft, Taipei, Taiwan
  }
}


\maketitle

\begin{abstract}
Hearing loss simulation models are essential for hearing aid deployment. However, existing models have high computational complexity and latency, which limits real-time applications, and lack direct integration with speech processing systems. To address these issues, we propose Neuro-MSBG, a lightweight end-to-end model with a personalized audiogram encoder for effective time-frequency modeling. Experiments show that Neuro-MSBG supports parallel inference and retains the intelligibility and perceptual quality of the original MSBG, with a Spearman’s rank correlation coefficient (SRCC) of 0.9247 for Short-Time Objective Intelligibility (STOI) and 0.8671 for Perceptual Evaluation of Speech Quality (PESQ). Neuro-MSBG reduces simulation runtime by 46 times (from 0.970 seconds to 0.021 seconds for a 1 second input), further demonstrating its efficiency and practicality.
\end{abstract}


\section{Introduction}
 Hearing loss simulation models aim to simulate how hearing impairment affects sound processing in the auditory system and have become essential tools in both research and evaluation. For example, the \textit{Clarity Challenge}~\cite{graetzer2021clarity} uses the Moore, Stone, Baer, and Glasberg (MSBG) model~\cite{Moore1997, baer1993spectral, baer1994interfering, moore1993loudness} to simulate individual perceptual conditions based on audiograms. Similarly, the \textit{Cadenza Challenge}~\cite{roa2023icassp_cadenza} and the \textit{Clarity Challenge} adopt perceptually grounded metrics such as the hearing-aid speech perception index (HASPI)~\cite{kates2014haspi}, the hearing-aid speech quality index (HASQI)~\cite{kates2010hasqi}, and the hearing-aid audio quality index (HAAQI)~\cite{kates2014haaqi}, which embed auditory processing to assess quality and intelligibility under hearing loss conditions.

Existing hearing loss models are generally divided into two categories: physiological models and engineering-oriented models. Physiological models, such as the model proposed in~\cite{Zilany2006} and the transmission-line (TL) cochlear model~\cite{Verhulst2018}, are designed to accurately model cochlear mechanics, but their complexity limits real-time integration. In contrast, engineering-oriented models, such as the Hohmann filterbank~\cite{Hohmann2002}, the Auditory Toolbox~\cite{Slaney1998}, and MSBG, balance deployment practicality with perceptual accuracy and  provide computational stability, making them well suited for real-time speech processing. Among engineering-oriented models, MSBG~\cite{Moore1997} is currently the most widely used. It simulates sensorineural hearing loss based on audiograms and reproduces key perceptual effects. Despite its widespread adoption in both academic and practical applications, MSBG has two major limitations:
(i) it does not support parallel processing, which reduces its applicability to real-time speech systems; and
(ii) it relies on multiple filtering stages, which introduces variable delays. These limitations restrict the integration of MSBG into end-to-end learning frameworks and reduce its effectiveness in real-time or large-scale speech processing.

Recent studies~\cite{VanDenBroucke2020, Irino2022} have attempted to simplify both physiological and engineering-oriented models for real-time applications. For example, CoNNear~\cite{VanDenBroucke2020} simplifies the TL cochlear model and supports real-time simulation of auditory nerve responses. Similarly, P Leer et al.~\cite{tan2024how} trained neural models to emulate the Verhulst auditory periphery model for varying hearing-loss profiles. However, the lack of waveform-level output generation prevents waveform-level supervision and reduces its applicability to speech processing and hearing aid systems. For engineering-oriented models, the Wakayama University Hearing Impairment Simulator (WHIS)~\cite{Irino2022} addresses the latency and computational cost issues associated with MSBG. WHIS first computes the cochlear excitation pattern of a target hearing-impaired individual using a Gammachirp filterbank, and then dynamically generates a time-varying minimum-phase filter based on this pattern to transform normal-hearing speech into its hearing-loss-simulated counterpart. This method reduces the processing time for one second of speech to approximately 10 milliseconds while maintaining near-perfect temporal alignment with the original waveform. Despite its efficiency, WHIS still relies on frame-by-frame computation of infinite impulse response (IIR) filter coefficients and dynamic gain selection, and has not been optimized for vectorized or parallel processing. Therefore, its integration into an end-to-end deep learning framework remains challenging, and joint optimization with compensation models remains an open research direction.


With the increasing use of differentiable hearing loss models in hearing aid compensation, optimizing their design and performance has become a research focus. In physiological modeling, the auditory nerve responses generated by CoNNear have been used as loss functions to guide the training of compensation models~\cite{drakopoulos2023neural, drakopoulos2022differentiable, drakopoulos2023dnn}. This approach attempts to make the neural responses of hearing-impaired people when receiving compensated speech similar to the neural responses of normal-hearing people when listening to the original signal. In engineering-oriented models, Tu et al. proposed the Differentiable Hearing Aid Speech Processing (DHASP) framework~\cite{tu2021dhasp}, which reimplements the auditory processing pipeline in HASPI~\cite{kates2014haspi} and uses differentiable modules for backpropagation. Tu et al. also introduced a differentiable version of the MSBG model and applied it to the training of hearing aid algorithms~\cite{tu2021optimising}. These engineering-oriented approaches typically consist of differentiable finite impulse response (FIR) filters and audio processing steps designed to approximate auditory mechanisms. 

\begin{figure*}[t]
  \centering
  \includegraphics[width=\textwidth]{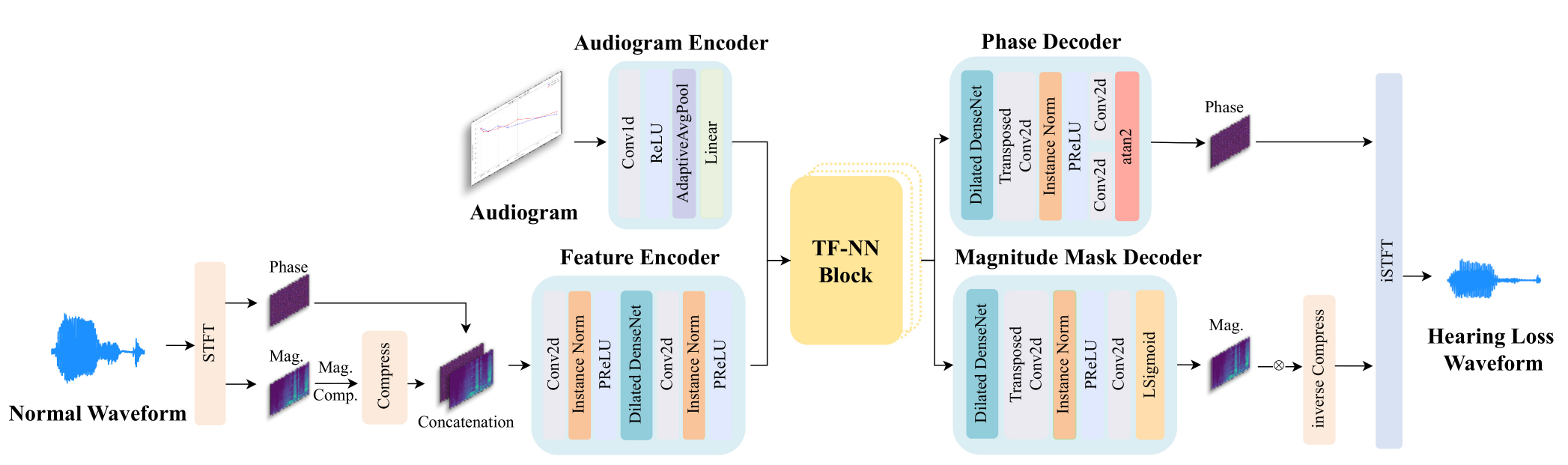}
  \caption{\textbf{Overview of the proposed Neuro-MSBG framework.}}
  \label{fig:overview}
\end{figure*}

Despite some progress in differentiable engineering-oriented hearing loss models, most efforts have focused on magnitude-domain simulation, with limited attention paid to the role of phase information. Meanwhile, recent advances in speech enhancement have highlighted the importance of phase modeling for perceptual quality. For instance, MP-SENet~\cite{lu2023mp} adopts a joint enhancement strategy for both magnitude and phase spectra, achieving significantly better performance than traditional magnitude-only methods and highlighting the importance of incorporating phase modeling. Inspired by these findings, we investigate the role of phase information in hearing loss simulation and propose Neuro-MSBG, an end-to-end fully differentiable hearing loss model. Neuro-MSBG outputs simulated audio in the waveform domain, which can be directly integrated with modern speech enhancement systems that rely on waveform-based losses and evaluation metrics (e.g., mean squared error (MSE), short-time objective intelligibility (STOI)~\cite{taal2011algorithm}, and perceptual evaluation of speech quality (PESQ)~\cite{rix2001perceptual}). It also supports noisy speech input, further enhancing its practical applicability. Experimental results show that the addition of phase processing significantly improves the fidelity of MSBG hearing loss simulation, highlighting the importance of phase modeling in replicating authentic auditory perception. The main contributions of our model are as follows:

\begin{itemize}[noitemsep, topsep=0pt]
    \item \textbf{Parallelizable and lightweight simulation:} Neuro-MSBG achieves parallel inference, reducing the simulation time for one second of audio from 0.970 seconds in the original MSBG to 0.021 seconds, a \textbf{46× speedup}.
    \item \textbf{Seamless integration with end-to-end speech systems:} By resolving the delay issues inherent in the original MSBG, Neuro-MSBG can be integrated into modern speech compensator training pipelines.
    \item \textbf{Phase-aware modeling:} By incorporating phase information, Neuro-MSBG maintains the intelligibility and perceptual quality of the original MSBG, achieving a Spearman’s rank correlation coefficient (SRCC) of 0.9247 for STOI and 0.8671 for PESQ.
\end{itemize}

At the end of this paper, we also demonstrate the preliminary integration of Neuro-MSBG into a speech compensator pipeline, thus confirming its practicality as a differentiable hearing loss simulation module. The remainder of this paper is organized as follows. Section II presents the proposed method. Section III describes the experimental setup and results. Finally, Section IV presents conclusions. 


\section{Methodology}
This section introduces the model architecture and training criteria of Neuro-MSBG. As shown in Fig. 1, the model takes normal speech signals and monaural audiograms as input.
The audiogram is transformed into personalized hearing features through the Audiogram Encoder, while the speech signal is converted into the time-frequency domain through Short-Time Fourier Transform (STFT) to obtain magnitude and phase features. These three types of features, including personalized hearing features, magnitude features, and phase features, are concatenated and then input into the Neural Network Block (NN Block).
The network then branches into two decoders: the Magnitude Mask Decoder and the Phase Decoder, which respectively predict the magnitude and phase shifts associated with hearing loss. Finally, the predicted magnitude and phase are combined to reconstruct the speech signal perceived by the hearing-impaired listener through inverse STFT.

\subsection{Neuro-MSBG}
Our model adopts an architecture based on MP-SENet~\cite{lu2023mp} and the advanced SE-Mamba framework~\cite{chao2024investigation}. This design is inspired by our experimental findings that phase information is critical for accurate hearing loss simulation (see Section III for details). To evaluate how well different neural modules capture spectral and temporal cues, we replace the original attention-based components with alternatives such as LSTM, Transformer, CNN, and Mamba blocks. Given Mamba’s efficiency in modeling long sequences and its low latency, we also control the number of parameters to ensure that the model remains lightweight and effective.

For audiogram integration, previous methods typically concatenate the audiogram representation along the frequency dimension. In contrast, we found that treating the audiogram as an additional input channel in addition to magnitude and phase produces more stable and effective results. This channel-based integration may enable the model to receive more consistent, spatially aligned conditioning across layers, thereby improving its ability to modulate internal feature representations.

\subsubsection{Audiogram Encoder}
To incorporate personalized hearing profiles, we design a lightweight Audiogram Encoder that transforms the audiogram $\mathbf{a} \in \mathrm{R}^{B \times 8}$ into a frequency-aligned representation $\mathbf{a}_{\text{enc}} \in \mathrm{R}^{B \times F}$, where $B$ denotes the batch size, and $F=201$ matches the STFT resolution. The transformation process is defined as:
\begin{equation}
\mathbf{a}_{\text{enc}} = \mathbf{W} \cdot \text{Flatten}\left( \text{AvgPool}\left( \sigma \left( \text{Conv}(\mathbf{a}) \right) \right) \right),
\label{eq:aud_enc}
\end{equation}
where Conv is a 1D convolution layer, $\sigma$ denotes a ReLU activation, and $\mathbf{W}$ is a linear projection matrix.

The encoded vector is then broadcast along the time axis and concatenated with the magnitude and phase features to form the three-channel input $\mathrm{R}^{B \times 3 \times T \times F}$ of the DenseEncoder and NN Block. This channel-based integration enables the model to consistently inject hearing-profile information in both time and frequency dimensions.

\subsubsection{Neural Network Blocks}
To capture the temporal and spectral structure of hearing-loss-affected speech, we design and compare multiple NN Blocks, each of which adopts a dual-path architecture to process the time and frequency dimensions separately. The input tensor is first rearranged and reshaped for temporal modeling, followed by a similar process for frequency modeling. Each path contains residual connections and a ConvTranspose1d layer to restore the original shape to ensure compatibility with subsequent modules.

In this unified framework, we replace the core time-frequency module with one of the following four alternatives:

\begin{itemize}
    \item \textbf{Mamba Block}: Combined with bidirectional Mamba modules, time and frequency are modeled separately to provide efficient long-range dependency modeling.
    \item \textbf{Transformer Block}: Transformer encoder layers are applied to both axes, and global attention is used to capture contextual information.
    \item \textbf{LSTM Block}: Bidirectional LSTM is used to model sequential patterns, and then linear projection is performed to maintain dimensional consistency.
    \item \textbf{CNN Block}: One-dimensional convolution is used to extract local features, followed by channel expansion, activation, and residual fusion.
\end{itemize}

This dual-axis design forms a flexible and stable framework for the simulation of hearing loss. It also allows for systematic comparison across architectures, demonstrating Mamba’s potential for low-latency, high-fidelity speech modeling.

\subsection{Training Criteria}

We adopt a multi-objective loss to jointly supervise spectral accuracy, phase consistency, and time-domain fidelity. The \textbf{magnitude loss} $\mathcal{L}_{\text{Mag}}$ is defined as the MSE between the predicted magnitude $\hat{m}$ and the ground-truth magnitude $m$:
\begin{equation}
\mathcal{L}_{\text{Mag}} = \frac{1}{N} \sum_{i=1}^{N} \left\| \hat{m}_i - m_i \right\|^2,
\label{eq:mag_loss}
\end{equation}
where $N$ is the number of training samples.
The \textbf{phase loss} $\mathcal{L}_{\text{Pha}}$ is inspired by the anti-wrapping strategy proposed in~\cite{ai2023neural} and consists of three components: the instantaneous phase loss $\mathcal{L}_{\mathrm{IP}}$, the group delay loss $\mathcal{L}_{\mathrm{GD}}$, and the integrated absolute frequency loss $\mathcal{L}_{\mathrm{IAF}}$, defined as:
\begin{align}
\mathcal{L}_{\mathrm{IP}}  &= \mathbb{E}_{p, \hat{p}} \left[ \left\| f_{\mathrm{aw}}\left(p - \hat{p}\right) \right\|_1 \right] \label{eq:lip}, \\
\mathcal{L}_{\mathrm{GD}}  &= \mathbb{E}_{p, \hat{p}} \left[ \left\| f_{\mathrm{aw}}\left( \Delta_{\mathrm{F}} (p - \hat{p}) \right) \right\|_1 \right] \label{eq:lgd}, \\
\mathcal{L}_{\mathrm{IAF}} &= \mathbb{E}_{p, \hat{p}} \left[ \left\| f_{\mathrm{aw}}\left( \Delta_{\mathrm{T}} (p - \hat{p}) \right) \right\|_1 \right], \label{eq:liaf}
\end{align}
\begin{equation}
\mathcal{L}_{\text{Pha}} = \mathcal{L}_{\text{IP}} + \mathcal{L}_{\text{GD}} + \mathcal{L}_{\text{IAF}},
\label{eq:pha_loss}
\end{equation}
where $f_{\text{aw}}(\cdot)$ denotes the anti-wrapping function used to mitigate $2\pi$ discontinuity, $\Delta_{\mathrm{F}}(\cdot)$ and $\Delta_{\mathrm{T}}(\cdot)$ represent the first-order differences of the phase error along the \textit{frequency} axis and \textit{time} axis, respectively.
The \textbf{complex loss} $\mathcal{L}_{\text{Com}}$ measures the MSE between the predicted complex spectrogram $\hat{c}$ and the ground-truth complex spectrogram $c$ (including both real and imaginary parts):
\begin{equation}
\mathcal{L}_{\text{Com}} = 2 \cdot \frac{1}{N} \sum_{i=1}^{N} \left\| \hat{c}_{i} - c_{i} \right\|^2.
\label{eq:com_loss}
\end{equation}
The \textbf{time-domain loss} $\mathcal{L}_{\text{Time}}$ is calculated as the $L_1$ distance between the predicted waveform $\hat{x}$ and the reference waveform $x$ to preserve temporal fidelity:
\begin{equation}
\mathcal{L}_{\text{Time}} = \frac{1}{N} \sum_{i=1}^{N} \left\| \hat{x}_{i} - x_{i} \right\|_1.
\label{eq:time_loss}
\end{equation}
The total training loss is a weighted sum of all components:
\begin{equation}
\mathcal{L}_{\text{Total}} = \lambda_{\text{Mag}} \mathcal{L}_{\text{Mag}} + \lambda_{\text{Pha}} \mathcal{L}_{\text{Pha}} + \lambda_{\text{Com}} \mathcal{L}_{\text{Com}} + \lambda_{\text{Time}} \mathcal{L}_{\text{Time}},
\label{eq:total_loss}
\end{equation}
where each $\lambda$ is a tunable scalar weight used to balance the contribution of each loss term.
\begingroup
\small
\setlength{\tabcolsep}{3pt}

\begin{table*}[t]
\centering
\caption{Performance comparison between monolithic models and our modular Neuro-MSBG framework with different TF-NN blocks.}

\label{tab:metric_comparison}
\begin{small}
\begin{tabular}{c|cccccccc}
\toprule
\textbf{Model Type} & \textbf{Architecture} & \textbf{STOI LCC} & \textbf{STOI SRCC} & \textbf{STOI MSE} & \textbf{PESQ LCC} & \textbf{PESQ SRCC} & \textbf{PESQ MSE} & \textbf{RAW MSE} \\
\midrule
\multirow{3}{*}{Monolithic}
& Transformer  &  0.7123  & 0.6932  & 0.0028  & 0.6026  & 0.6250  & 0.1677  & 0.0870  \\
& CNN         & 0.8156  & 0.8037  & 0.0019  & 0.6058  & 0.6433  & 0.1455  & \underline{0.0670}  \\
& LSTM          & 0.5004 & 0.5064 & 0.0137 & 0.3291     & 0.4193     & 0.2242     & 0.1054 \\
\midrule
\multirow{4}{*}{\makecell[c]{\textbf{Neuro-MSBG}\\[-0.05ex]{\footnotesize\mbox{TF-NN Block Replacement}}}}
 & Transformer & 0.7271 & 0.7593 & 0.0012 & 0.5326 & 0.5823 & 0.4910 & 0.0529 \\
 & CNN          & 0.8234  & 0.8591  &  0.0009  & 0.7374  & 0.7580  & 0.1606  & \underline{0.0670}  \\
 & \underline{LSTM}        & \underline{0.8443} & \underline{0.8999} & \underline{0.0006} & \underline{0.8226} & \underline{0.8312} & \underline{0.0801} & 0.0691 \\
 & \textbf{Mamba} & \textbf{0.8475} & \textbf{0.9247} & \textbf{0.0006} & \textbf{0.8519} & \textbf{0.8671} & \textbf{0.0782} & \textbf{0.0669} \\
\bottomrule

\end{tabular}
\end{small}
\end{table*}
\endgroup

\section{Experiment}
The dataset comprises 12,396 clean utterances (11,572 for training and 824 for testing) from VoiceBank~\cite{Veaux2013VoiceBank} and their noisy counterparts from VoiceBank–DEMAND~\cite{valentini2017noisy}, created by mixing each clean utterance with one randomly selected noise type from 10 DEMAND recordings~\cite{Thiemann2013DEMAND} and one signal-to-noise-ratio (SNR) level. To ensure generalization, 8 noise types are used for training and 2 disjoint types for testing, with SNR levels drawn from \{0, 5, 10, 15\} dB  for training and \{2.5, 7.5, 12.5, 17.5\} dB  for testing. Each utterance is paired with two monaural audiograms, representing different levels of hearing loss, ranging from mild to severe.  
Consequently, the training set is expanded to
\(11,572 \times 2 \times 2 = 46,288\) samples, and the test set is expanded to
\(824 \times 2 \times 2 = 3,296 \) samples.  
Speech data and audiograms are disjoint across training and testing splits, ensuring that the model is evaluated on unseen utterances and unseen hearing-loss profiles.

Specifically, this study conducted a series of experiments covering five main aspects: (i) comparison of Neuro-MSBG with different architectures and monolithic baselines (Table~\ref{tab:metric_comparison}); (ii) runtime evaluation (Table~\ref{tab:inference_time}); (iii) validation of the necessity of phase prediction (Table~\ref{tab:mag_only}); (iv) comparison of audiogram integration strategies (Table~\ref{tab:aud_enc_ablation}); and (v) application of Neuro-MSBG in a speech compensator (Table~\ref{tab:haspi_hasqi}). In all quantitative tables, \textbf{bold} highlights the best-performing model and \underline{underline} marks the second best.
All experiments were performed on a single NVIDIA RTX 3090. The models were trained for 200 epochs with a batch size of 3, an initial learning rate of 0.0005, and the AdamW optimizer.

\subsection{Data Preparation}
We use the MSBG model to simulate hearing loss by applying an ear-specific gain curve to each input, resulting in single-channel impaired speech. However, due to the multi-stage filtering in MSBG, the output may exhibit unpredictable delay relative to the original signal. Such misalignment is undesirable for downstream tasks, such as speech enhancement or intelligibility assessment.

To estimate and correct this delay, we generate an auxiliary reference signal along with the main audio during the MSBG process~\cite{graetzer2021clarity}. 
This auxiliary signal is a silent waveform with the same length and sampling rate as the input, used solely for delay tracking.  
A unit impulse is inserted into this signal, which is defined as:
\begin{equation}
\delta[n - k] =
\begin{cases}
1, & \text{if } n = k \\
0, & \text{otherwise}
\end{cases}, 
\end{equation}
where \( n \) denotes the discrete time index, and \( k = \frac{F_s}{2} \) is the sample position of the impulse, where \( F_s \) represents the sampling rate in Hz.

\begin{figure}[t]
    \centering
    \includegraphics[width=0.9\linewidth]{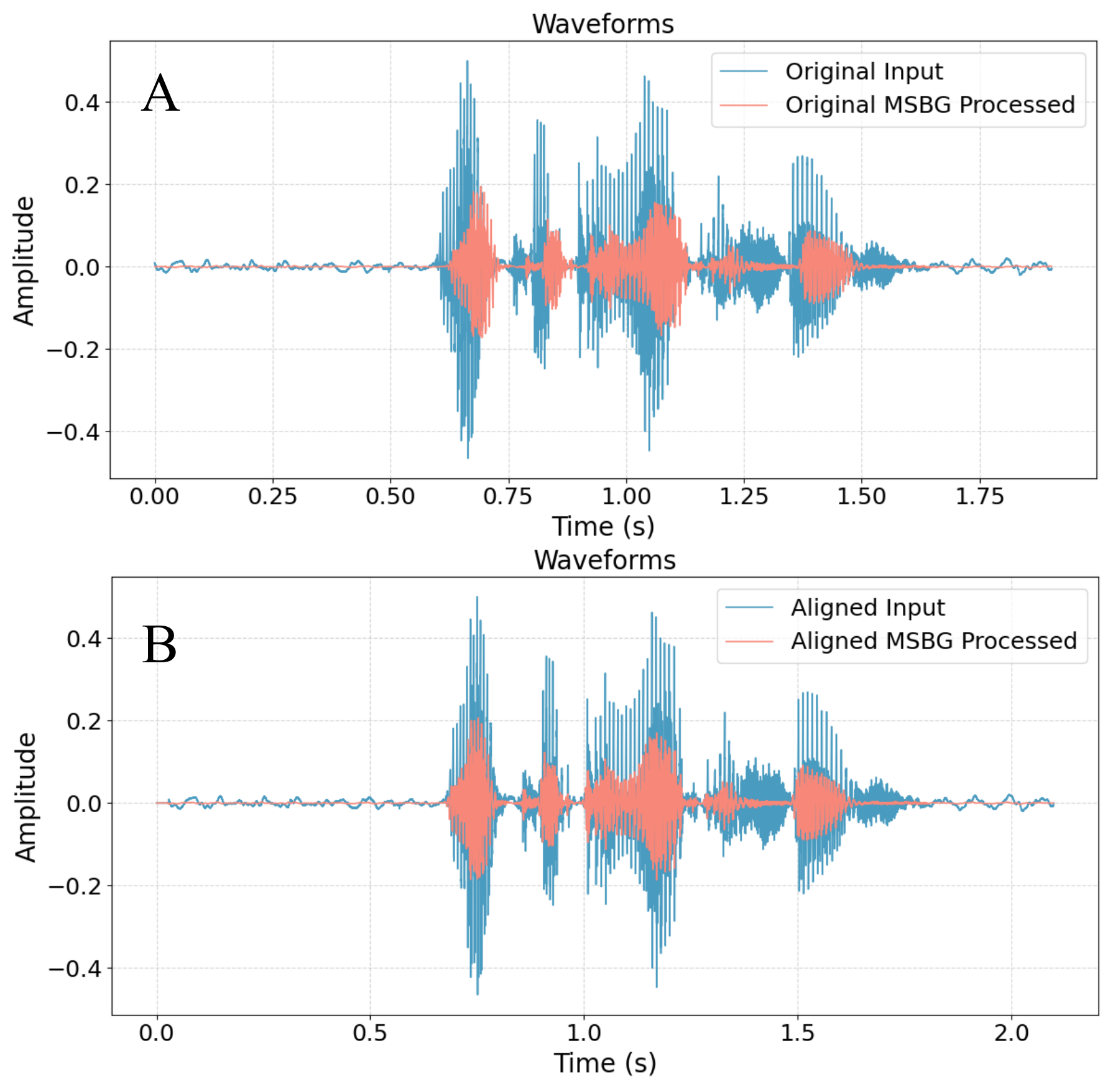}
    \caption{\textbf{Waveform alignment before and after shifting.} In A, the MSBG-processed signal (red) shows a clear delay relative to the original input (blue). In B, the waveforms are time-aligned using impulse-based method, allowing for a fair and accurate assessment of the effect of MSBG.}
    \label{fig:aligned_waveform}
\end{figure}

The impulse is inserted at the midpoint of the auxiliary reference signal. After MSBG simulation, we compare the pre/post impulse positions to estimate the delay introduced by MSBG. We then use the estimated delay to time-align the original normal-hearing input, the clean reference, and the impaired output to ensure accurate evaluation through metrics such as STOI and PESQ that are highly sensitive to timing errors.
Each training instance consists of: 
1) the single-ear impaired speech,  
2) the aligned clean reference,
3) the aligned normal-hearing input, and 4) the associated 8-dimensional audiogram vector. Fig.~\ref{fig:aligned_waveform} shows the waveform alignment before and after the delay‐compensation shift.

\subsection{Experimental Results}
In the early stages of our experiment, we used monolithic models with unified architectures such as CNN, LSTM, and Transformer for hearing simulation. However, due to limited performance, we subsequently adopted the Neuro-MSBG framework, replacing only the TF-NN block with different architectures, including CNN, LSTM, Transformer, and Mamba. The results are shown in Table~\ref{tab:metric_comparison}. To ensure a fair comparison across different architectures, the number of parameters of all models was set to be roughly the same. From Table~\ref{tab:metric_comparison}, we observe that the Neuro-MSBG variants consistently outperform the monolithic baselines on unseen test data. Among them, Neuro-MSBG with Mamba achieves the best performance across all evaluation metrics.

Next, we compare Neuro-MSBG with an existing neural network-based model,  CoNNear, in terms of model size and inference time. Although the two models have different goals—CoNNear simulates physiologically grounded auditory nerve responses, while our framework focuses on perceptual signal transformation—the comparison is appropriate given their common goal of real-time hearing loss modeling. As shown in Table~\ref{tab:inference_time}, CoNNear has approximately 11.7 million parameters, while Neuro-MSBG has only 1.45 million parameters. In addition to its lightweight architecture, Neuro-MSBG supports noisy input conditions and accommodates a wide range of audiogram configurations, providing additional advantages for practical applications involving diverse acoustic environments and personalized hearing loss profiles.

Table~\ref{tab:inference_time} also shows the inference time of different models. MSBG does not support parallel processing and cannot be executed on GPU; therefore, the corresponding GPU column is marked as NA. In contrast, Neuro-MSBG (Mamba) leverages a CUDA-accelerated selective scan kernel for core operations, which currently only supports GPU execution. Therefore, only the inference time on GPU is reported. In terms of inference time, Neuro-MSBG (Mamba) achieves about 46$\times$ speedup on GPU over MSBG on CPU. For CPU-executable variants such as Neuro-MSBG (LSTM), the inference time is 0.016 seconds, which is 60$\times$ faster than MSBG’s 0.970 seconds. In addition, we also implemented and evaluated the CoNNear model. Despite its larger parameter size (11.7 million), it shows fast inference in our computation environment, with a GPU runtime of 0.099 seconds and a notably fast CPU runtime of 0.025 seconds. The faster CPU runtime of CoNNear than the GPU version is likely due to the batch size of 1 used in this experiment, which limits the benefits of GPU parallelism. 

\begingroup
\scriptsize
\begin{table}[t]
\centering
\caption{Comparison of runtime of MSBG and Neuro-MSBG on different devices. We measured the inference time required to process a 1-second, 44.1 kHz audio signal using an Intel Xeon Gold 6152 CPU and an NVIDIA RTX 3090 GPU.}
\label{tab:inference_time}
\begin{tabular}{lcccc}
\toprule
\textbf{Model} & \textbf{CPU} & \textbf{GPU}  & \textbf{Param}  \\
\midrule
MSBG        & 0.970 & NA      &  NA   \\
Neuro-MSBG (Mamba)  &  NA & 0.021 s       & 1.45M \\
Neuro-MSBG (LSTM)  & 0.617 s & 0.016 s & 1.47M \\
Neuro-MSBG (CNN)  & 0.592 s & 0.016 s & 1.45M \\
CoNNear  & 0.025 s & 0.099 s   &  11.7M \\
\bottomrule
\end{tabular}
\end{table}
\endgroup
\subsection{Ablation Study}
Previous approaches to modeling hearing loss typically predict only the magnitude spectrum while reusing the input phase for waveform reconstruction. We initially adopted this approach; however, our empirical analysis revealed that phase information plays a crucial role in MSBG-based simulation. 
We conducted an ablation study using Neuro-MSBG (Mamba). 
As shown in Table~\ref{tab:mag_only}, predicting both magnitude and phase substantially outperforms magnitude-only prediction in all metrics (STOI MSE, PESQ MSE, and waveform-level MSE). Specifically, the STOI MSE decreased from 0.0041 to 0.0006, indicating a notable improvement in intelligibility, while the PESQ MSE decreased from 2.3579 to 0.0782, reflecting improved perceptual quality. At the waveform level, the MSE decreased from 0.0986 to 0.0669, confirming that phase modeling is critical for both perceptual fidelity and accurate signal reconstruction. 

Furthermore, in many speech-related applications involving audiograms, a common practice is to concatenate the audiogram vector with the audio features before feeding them into the model. We initially adopted this simple approach, but found that it could not effectively capture the relationship between hearing profiles and spectral features. To address this issue, we introduced a lightweight Audiogram Encoder that transforms the 8-dimensional audiogram vector into a frequency-aligned representation. This representation is appended as a third channel along with the magnitude and phase features. As shown in Table~\ref{tab:aud_enc_ablation}, incorporating the Audiogram Encoder leads to consistent reduction in STOI MSE, PESQ MSE, and waveform-level MSE, demonstrating its effectiveness in integrating personalized hearing profiles to improve hearing loss modeling.

\begingroup
\tiny
\begin{table}[t]
\centering
\caption{Performance comparison of NEURO-MSBG models with and without phase prediction.}
\label{tab:mag_only}
\begin{tabular}{lccc}
\toprule
\textbf{Setting} & \textbf{STOI MSE} & \textbf{PESQ MSE} & \textbf{RAW MSE} \\
\midrule
Magnitude Only    & 0.0041 & 2.3579 & 0.0986 \\
Magnitude + Phase & \textbf{0.0006} & \textbf{0.0782} & \textbf{0.0669} \\
\bottomrule
\end{tabular}
\end{table}
\endgroup

\subsection{Qualitative Evaluation}
\begin{figure}[t]
    \centering
    \includegraphics[width=1\linewidth]{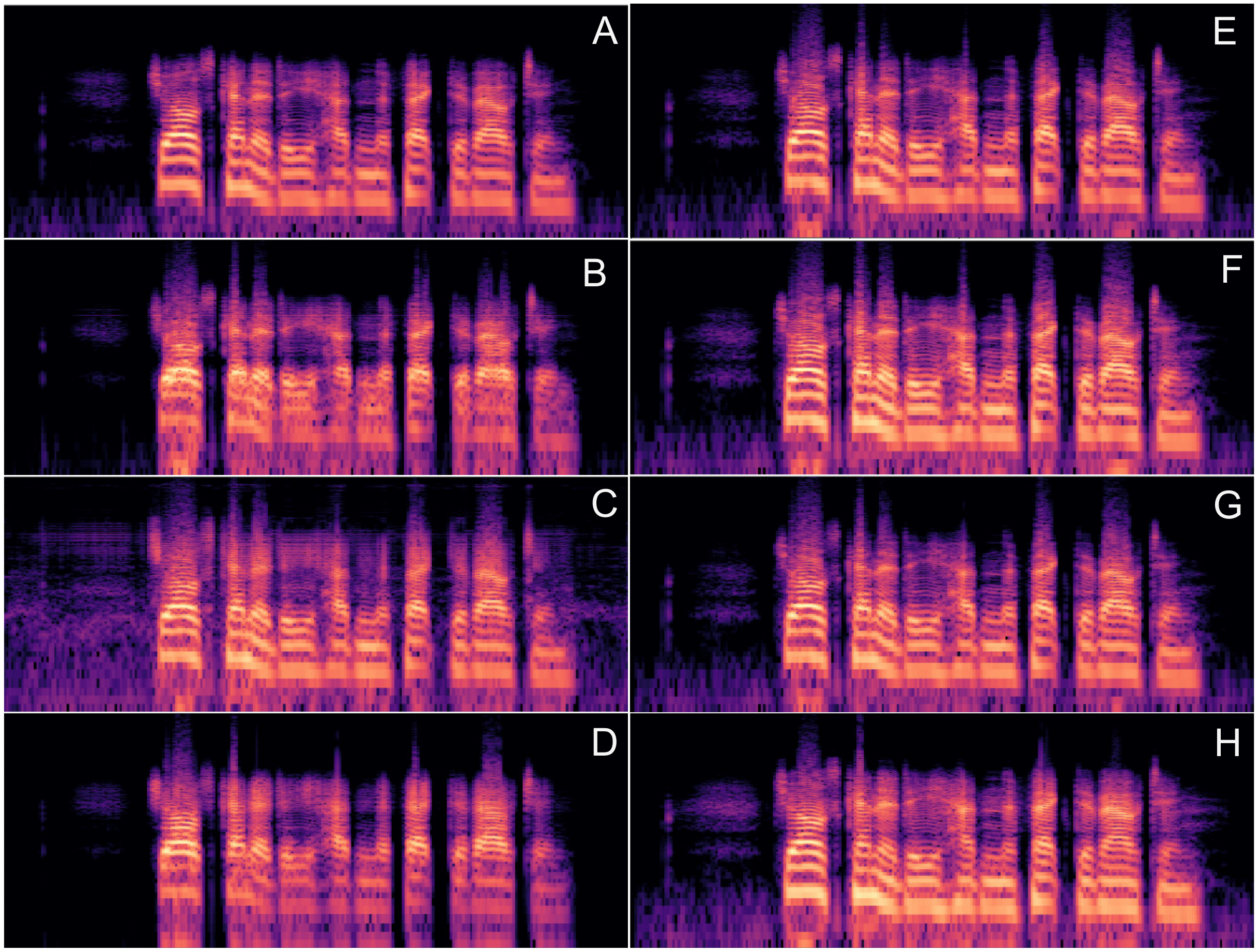}
    \caption{
    Log-magnitude spectrograms of speech outputs of seven models and ground-truth speech: (A) ground truth, (B), (C), and (D) outputs of baseline CNN, LSTM, and Transformer models, and (E), (F), (G), and (H) outputs of Neuro-MSBG models using Mamba, CNN, LSTM, and Transformer Blocks.
    }
    \label{fig:qualitative}
    \vspace{-5pt}
\end{figure}
To evaluate the model's performance in reconstructing hearing-loss-affected speech, we compare the log-magnitude spectrograms of speech outputs of seven models and ground-truth speech (Fig.~\ref{fig:qualitative}). Among them, Neuro-MSBG (Mamba) produces the most accurate reconstruction, preserving the harmonic structure and high-frequency energy. Neuro-MSBG with CNN, LSTM, and Transformer Blocks also preserve key spectral features but exhibit some energy imbalance or mild distortion. In contrast, the baseline models introduce more obvious artifacts: the CNN and LSTM variants lack clarity and high-frequency content, while the Transformer variant has difficulty reconstructing an accurate spectrogram.

\begingroup
\tiny
\begin{table}[t]
\centering
\caption{Effects of the Audiogram Encoder in the Neuro-MSBG model.}
\label{tab:aud_enc_ablation}
\begin{tabular}{lccc}
\toprule
\textbf{Setting} & \textbf{STOI MSE} & \textbf{PESQ MSE} & \textbf{RAW MSE} \\
\midrule
w/o Audiogram Encoder & 0.0013 & 0.1152 & 0.0670 \\
w/ Audiogram Encoder & \textbf{0.0006} & \textbf{0.0782} & \textbf{0.0669} \\
\bottomrule
\vspace{-5pt}
\end{tabular}
\end{table}
\endgroup

\subsection{Training a Compensator with Neuro-MSBG}
To advance end-to-end hearing loss compensation, recent work (e.g., NeuroAMP~\cite{Ahmed2024NeuroAMP}) has integrated audiogram-aware processing directly into neural networks, replacing traditional modular pipelines with personalized, data-driven amplification. Inspired by this direction, we propose a complementary approach that connects a trainable compensator to a frozen, perceptually grounded simulator (Neuro-MSBG), enabling the compensator to shape its input to match the individual's hearing profiles.

\begin{figure}[t]
    \centering
    \includegraphics[width=1\linewidth]{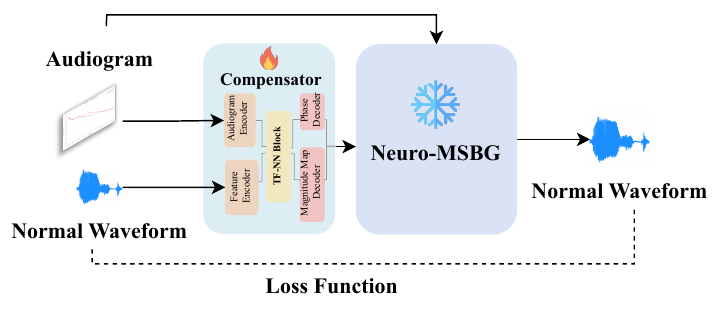}
    \caption{\textbf{Training framework of a personalized speech compensator using Neuro-MSBG as a fixed hearing loss simulator.} The left module (trainable, labeled ``Compensator'') is optimized to minimize the loss between the Neuro-MSBG output and the original clean waveform. The right module (``Neuro-MSBG'') is frozen during training and used only for perceptual feedback.}
    \label{fig:speech_enhancement}
\end{figure}

Neuro-MSBG is lightweight, differentiable, and does not require clean reference alignment, making it suitable for integration into an end-to-end hearing loss compensation system. Building on this feature, we explore a new use case: connecting a pre-trained Neuro-MSBG model to a trainable compensator to achieve personalized hearing enhancement, as illustrated in Fig.~\ref{fig:speech_enhancement}. The training and test sets are from VoiceBank~\cite{Veaux2013VoiceBank}.

The goal of the compensator is to transform an input waveform into a personalized, compensated version. This compensated waveform is then passed into the frozen Neuro-MSBG model, with the training objective of closely matching the final output with the original clean speech. This design enables the compensator to function as a personalized module that adjusts the audio to each user's hearing condition. It is important to note that we did not fine-tune Neuro-MSBG, as our main objective was to initially verify the feasibility and effectiveness of integrating Neuro-MSBG into the training pipeline. The compensator adopts the same architecture as Neuro-MSBG, but with a key modification in the magnitude path: the original masking-based magnitude encoder is replaced by a mapping strategy designed to enhance or restore lost information. This adjustment better aligns the model with the goal of compensation, enabling the model to generate gain-adjusted outputs that enhance speech intelligibility for hearing-impaired users.

We evaluate the effectiveness of the proposed compensator using the HASPI metric to assess the improvement in perceptual speech intelligibility. As shown in Table~\ref{tab:haspi_hasqi}, the compensator significantly improves the average HASPI score from 0.428 to 0.616 ($\Delta = +0.187$). This improvement is statistically significant, supported by both the paired \textit{t}-test ($t = -24.113$, $p < 0.00001$) and the Wilcoxon signed-rank test ($W = 292426.0$, $p < 0.00001$). The observed effect size is moderately large ($d = 0.594$), indicating a substantial improvement in perceptual quality across samples.

\begin{table}[t]
\centering
\caption{Statistical comparison of HASPI scores before and after applying the proposed compensator.}
\label{tab:haspi_hasqi}
\resizebox{\columnwidth}{!}{
\begin{tabular}{lccc}
\toprule
\textbf{Metric} & \textbf{Original} & \textbf{Compensated} & \textbf{Change ($\Delta$)} \\
\midrule
HASPI (mean $\pm$ std) & 0.428 $\pm$ 0.360 & 0.616 $\pm$ 0.352 & $+0.187$ \\
\midrule
\textit{t-test (paired)} & \multicolumn{3}{c}{$t = -24.113$, $p < 0.00001$} \\
\textit{Wilcoxon signed-rank} & \multicolumn{3}{c}{$W = 292426.0 $, $p < 0.00001$} \\
\textit{Cohen's $d$ (paired)} & \multicolumn{3}{c}{$d = 0.594$} \\
\bottomrule
\vspace{-10pt}
\end{tabular}
}
\end{table}

\section{Conclusion}
\label{sec:typestyle}
This paper introduces Neuro-MSBG, a lightweight and fully differentiable hearing loss simulation model that addresses key limitations of traditional approaches, including delay issues, limited integration flexibility, and lack of parallel processing capabilities. Unlike conventional models that require clean reference signal alignment, Neuro-MSBG can be seamlessly integrated into end-to-end training pipelines and avoids timing mismatches that may affect evaluation metrics such as STOI and PESQ. Its parallelizable architecture and low-latency design make it well suited for scalable speech processing applications. In particular, the Mamba-based Neuro-MSBG achieves 46× speedup over the original MSBG, reducing the simulation time for one second of audio from 0.970 seconds to 0.021 seconds through parallel inference. Meanwhile, the LSTM-based variant achieves an inference time of 0.016 seconds, which is 60× faster than MSBG. Experimental results further demonstrate that jointly predicting magnitude and phase can significantly improve speech intelligibility and perceptual quality, with SRCC of 0.9247 for STOI and 0.8671 for PESQ. In addition, the proposed Audiogram Encoder can effectively transform audiogram vectors into frequency-aligned features, outperforming the simple concatenation method and more accurately modeling individual hearing profiles.


\end{document}